# A ductility metric for refractory-based multi-principal-element alloys

Prashant Singh,[a,*] Brent Vela,[b] Gaoyuan Ouyang,[a] Nicolas Argibay,[a] Jun Cui,[a,c] Raymundo Arroyave,[b] Duane D. Johnson,[a,c]

[a]Ames Laboratory, U.S. Department of Energy, Iowa State University, Ames, IA 50011, USA
[b]Department of Materials Science & Engineering, Texas A&M University, College Station, TX 77843, USA
[c]Department of Materials Science & Engineering, Iowa State University, Ames, IA 50011, USA

Corresponding author Email: psingh84@ameslab.gov/prashant40179@gmail.com

## Abstract

We propose a quantum-mechanical dimensionless metric, the local-lattice distortion (LLD), as a reliable predictor of ductility in refractory multi-principal-element alloys (RMPEAs). The LLD metric is based on electronegativity differences in localized chemical environments and combines atomic-scale displacements due to local lattice distortions with a weighted average of valence-electron count. To evaluate the effectiveness of this metric, we examined body-centered cubic (bcc) refractory alloys that exhibit ductile-to-brittle behavior. Our findings demonstrate that local-charge behavior can be tuned via composition to enhance ductility in RMPEAs. With finite-sized cell effects eliminated, the LLD metric accurately predicted the ductility of arbitrary alloys based on tensile-elongation experiments. To validate further, we qualitatively evaluated the ductility of two refractory RMPEAs, i.e., NbTaMoW and $Mo_{72}W_{13}Ta_{10}Ti_{2.5}Zr_{2.5}$, through the observation of crack formation under indentation, again showing excellent agreement with LLD predictions. A comparative study of three refractory alloys provides further insights into the electronic-structure origin of ductility in refractory RMPEAs. This proposed metric enables rapid and accurate assessment of ductility behavior in the vast RMPEA composition space.

**Keywords:** Ductility, Multi-principal-element alloys, DFT, CALPHAD, Experiment

---

## Introduction

Since the concept of multi-principal-element alloys (MPEAs) was proposed [**1,2**], numerous systems have been developed to enhance high-temperature phase stability, expanding the design prospects for new alloys [**3-22**]. Refractory MPEAs (RMPEAs) are a relatively new class of single-phase materials based on body-centered cubic (bcc) refractory elements often mixed with low-density bcc metals [**23,24**]. These alloys have received more attention than other metallic alloys due to their attractive properties, such high melting temperature, and a weak temperature-dependent yield strength, which is about 400 MPa near 1600°C [**25,26**]. However, RMPEAs generally have low ductility, even in compression. This brittle behavior is intrinsic to bcc metals, also exhibited in bcc RMPEAs [**27,28**]. Moreover, while there is a simple predictive metric for strength of any metal [**29**], no predictive correlation has yet been established between ductility and strength for these alloys. For example, the uniaxial yield strength is high in the refractory alloys, NbTaMoW (1 GPa), MoNbTaTi (1.2 GPa), NbTaTiW (1.8 GPa), and MoWNbTaV (1.25 GPa), though their ductility is in all cases low (elongation strain < 4%) [**30,31**]. Like most refractory elements, RMPEAs are quite brittle, with a relatively sharp brittle-to-ductile transition as temperature increases.

The strengthening mechanisms in RMPEAs have been the subject of extensive research. One widely recognized mechanism is solid-solution strengthening, which is influenced by local-lattice distortions (LLDs). This phenomenon contributes to the high strength observed in these alloys. Additionally, a disparity in atomic sizes and elastic moduli are thought to impede dislocation motion, which is a conventional strengthening mechanism. [**32-36**]. Several theoretical studies have provided some understanding of uniaxial yield strengths, e.g., [**37,38**]. However, the literature on approaches to predict ductility in RMPEAs remains sparse [**31,39**]. Lilensten *et al.* [**40**] and Huang *et al.* [**41**] proposed the idea of metastability engineering to improve the ductility in bcc RMPEAs. This idea was utilized for steels [**42,43**] and for Ti-based bcc [**44,45**] and fcc [**46**] MPEAs to enhance uniform tensile ductility [**47**].

In relation to the inherent ductility of the bcc lattice, researchers have proposed three readily available and widely accepted ductility indicators for RMPEAs: Pugh ratio [**48**], Cauchy pressure [**49**], and valence-electron concentration (VEC) [**50**]. Pugh's ratio (B/G) is a measure of a material's ductility, reflecting the competition between plastic deformation (shear modulus, G) and fracture strength (crack formation, represented by bulk modulus B). As such, the Pugh ratio provides a measure of the favorability of cracking vs. slip [**48**]. The Cauchy pressure (a difference between elastic constants, $C_{12} – C_{44}$) was proposed as a ductility indicator by Pettifor [**49**]. The Cauchy pressure is the difference between two elastic constants, $C_{12}$ - $C_{44}$, where a positive value indicates non-directional metallic bonds, and likely to have intrinsic ductility. Qi and Chrzan [**51**] proposed that the intrinsic ductility of a bcc refractory alloy can be estimated based on VEC. In addition, there have been attempts to design more nuanced ductility metrics, e.g., Hu et al. [**52**] created surrogate models for the Rice criterion [**53**] and demonstrated a limited correlation for a small dataset of fracture strain in RMPEAs.

While these metrics have been used to try to design intrinsically ductile RMPEAs when compared against compressive fracture strain, the results (compiled by Hu et al. [**52**]) demonstrated weak correlations with experimental values, see **Fig. 1**. The electronic origins of mechanical properties, such as ductility, in concentrated refractory alloys, including RMPEAs, remain poorly understood. An improved understanding of these underlying mechanisms would significantly accelerate the discovery of new and optimized alloys. Therefore, this study seeks new insights and a framework for accurately predicting ductility in chemically disordered concentrated refractory alloys, focusing on RMPEAs.

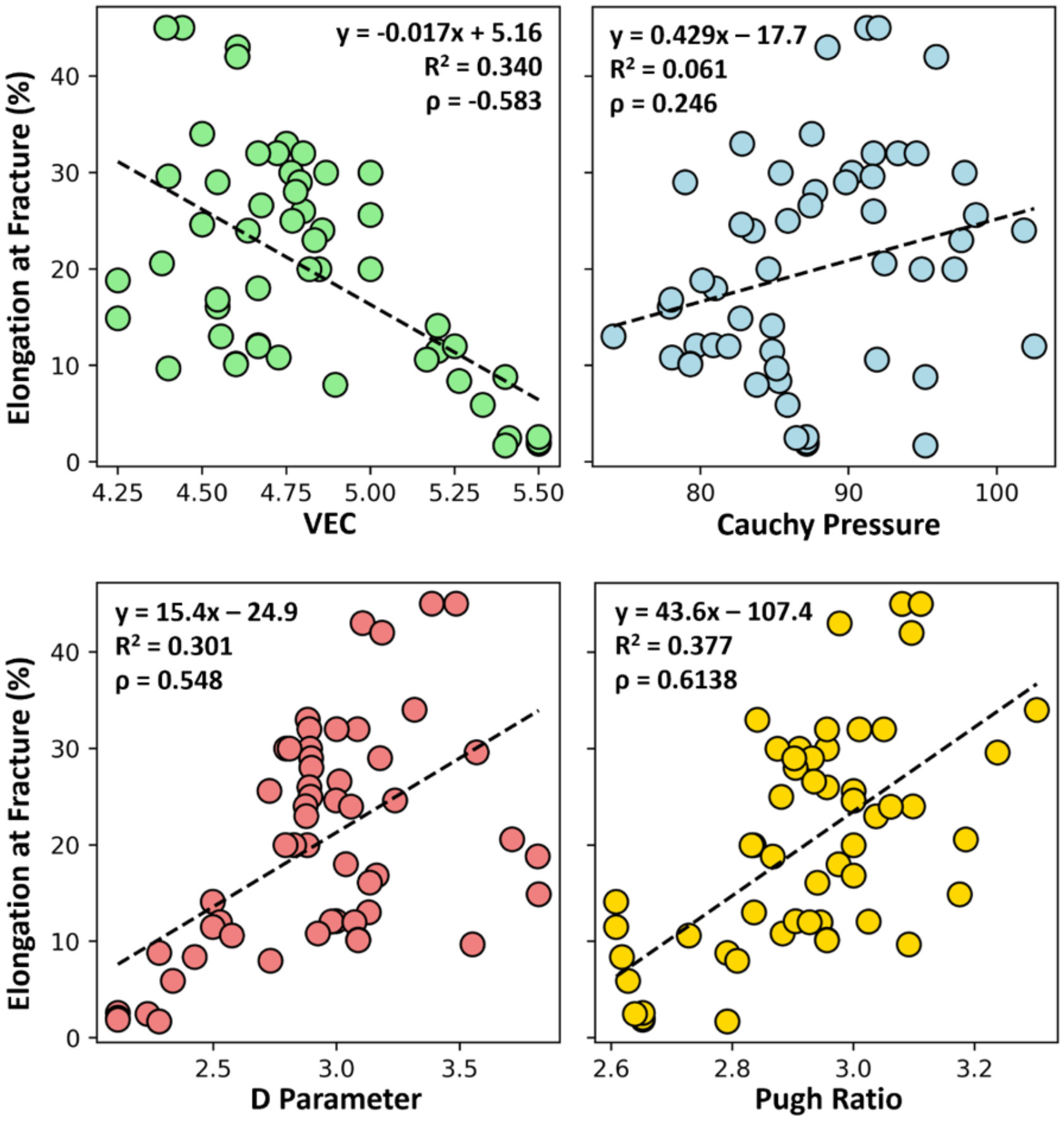

**Figure 1**. Linear models, coefficient of determination, and Pearson correlation between compressive fracture strain of 56 RMPEAs (see more detail in **Table A1**) and the following ductility indicators: VEC, Cauchy pressure, Pugh ratio, and the Rice criterion.

We hypothesize that the ductility observed in bcc RMPEAs can be attributed to quantum-mechanical phenomena linked to intrinsic characteristics, such as lattice distortions in the local chemical environments and chemical disorder, which impacts electronic band dispersion ("band structure"), exhibiting disorder broadening [**17**,**54**], in contrast to ordered alloys. We anticipate that the transition from ductile-to-brittle behavior in these alloys is closely tied to nanoscale structural features. We propose a dimensionless metric for bcc chemically random alloys that predicts ductility via quantities (average and $L_{2,1}$ norm) derived from atomic displacements ($\Delta u, \sqrt{[\Delta u]^2}$) relative to equilibrium atomic positions, obtained by minimizing Hellmann-Feynman forces using density-functional theory (DFT) calculations within a Super-Cell Random APproximates (SCRAPs) that mimic appropriate configurations for random alloys. A SCRAP is constructed as an optimal supercell of fixed size (number of sites) with the thermodynamically averaged (observable) atomic short-range order (here, we focus on homogeneous random alloys, where pair correlations are optimized to zero to the third-neighbor shell around every site). From each relaxed supercell, we extract displacements and derived quantities to construct a dimensionless 'LLD' metric to successfully characterize ductility in bcc RMPEAs and confirm results by experimental validation. The proposed LLD metric was also compared with the Rice-Thomson ductility criterion [**55**] and elongation to tensile strain from experiments to understand the correlation among the quantities of diverse origin. The trend in room-temperature (RT) yield-strength data was also assessed with respect to the LLD metric. The electronic structures (bonding, charge transfer, density of states, …) of a selected set of RMPEAs from ductile and brittle regions were investigated to understand its relationship with ductility (and its electronic origin). These new insights will guide our efforts to identify RMPEAs with improved RT ductility.

**Methods**

Density functional theory method: The total-energy calculations were performed using DFT methods, embodied in the Vienna *ab initio* simulation package (VASP) [**56**] plane-wave pseudo-potential method with projector-augmented waves (PAW) [**57**,**58**]. The Perdew-Burke Ernzerhof (PBE) [**59**] DFT exchange-correlation functional is used for non-spin-polarized generalized gradient approximation (GGA). The kinetic energy cutoff of 500 eV is employed for the plane-wave basis. The Monkhorst-Pack scheme for Brillouin zone integration [**60**] was carried out using 5×5×5 k-points meshes. The convergence threshold for energy is $10^{-5}$ eV, while the symmetry-unrestricted optimization for the geometry is performed using conjugate-gradient method until residual forces on each atom is below 0.01 eV/Å. A Super-Cell Random Approximates (SCRAPs) supercell of a 60-atom (5×3×2) and 90-atom (5×3×3) were used to mimic the homogeneously random RMPEAs, as generated using a Hybrid Cuckoo Search optimization method [**61**]. The in-depth understanding of the local lattice distortion and its correlation with ductility is expected to advance the design of RMPEAs for high-temperature applications. For equiatomic cases, we achieved correct correlations and spatial distribution within 60 and 90-atom SCRAPs. In particular, we evaluated the LLD metric for supercell sizes of stoichiometric quaternary NbTaTiV (i.e., i.e., 16 (2 × 2 × 2), 32 (4 × 2 × 2), 60 (5 × 3 × 2), 72 (4 × 3 × 3), 128 (4 × 4 × 4), 160 (5 × 4 × 4) atoms per supercell) and found that finite-cell effects beyond 60-atom are insignificant. However, for non-stoichiometric cases, 120-atom SCRAPs were needed to achieve proper atomic pair-correlation functions (to 3 neighbor shells) or spatial distribution. Beyond 120-atom SCRAPs, no noticeable changes in energy or LLD were found [**62**].

*Slab model generation and energy calculation*: A [110]-oriented bcc slab was generated with a 10Å vacuum, where the atoms are shifted along **c** direction for symmetric termination [**63,64**]. All slabs are constrained to have symmetric top and bottom surfaces. The surface energy ($\gamma_{\mathrm{hkl}}$) of bcc slabs for the facet with Miller index (110) was calculated using the expression:

$$\gamma_{hkl} = \frac{E_{hkl}^{slab} - n^{slab} \times E_{hkl}^{bulk}}{2 \times A^{slab}}, \qquad (1)$$

where $E_{hkl}^{slab}$ is the total slab energy, $E_{hkl}^{bulk}$ energy-per-atom of the conventional unit-cell, $n^{slab}$ is the number of atoms in the slab, and $A^{slab}$ is the surface area of the slab.

*CALPHAD for Novel RMPEA Design*: In addition to datasets from literature [**52**], a set of 17 RMPEAs designed throughout ARPA-E's ULTIMATE program are studied here [**65**], which are designed via a composition-agnostic, multi-constraint factorial exploration similar to the scheme used in our previous work [**66**]. However, this work has a larger initial design space consisting of every quinary, quaternary, and ternary combination of the 10 elements (Al, Cr, Hf, Mo, Nb, Ta, Ti, V, W, Zr). Each ternary, quinary, and quaternary system was sampled at 5 at.% resulting in 17 candidate RMPEAs. The downselected alloys were required to meet minimum constraints from Department of Energy's ARPA-E's ULTIMATE program as listed in **Table 1** [**65**]. Thermo-Calc's TCHEA5 CALPHAD database was used as information source to perform equilibrium simulations and down-select final set of RMPEAs. In addition, the Maresca-Curtin model for yield strength **[45]** was used to filter for alloys that likely have a $\sigma_{YS}$ > 200 MPa at 1300°C. The Maresca-Curtin model assumes the thermally activated glide of edge dislocations through randomly solute fields is the dominant strengthening mechanism in alloys with high lattice distortion. Inputs to this model, such as lattice parameters, elastic constants, and solute misfit volumes, are approximated with the rule-of-mixture, as used by the authors of the model.

**Table 1**. Summary of constraints and the information source associated with each constraint.

| **Constraint** | **Information Souce** |
|---|---|
| $\rho < 11$ g/cc at RT | Thermo-Calc Equilibrium Simulation |
| $T_{solidus} > 1500$°C | Thermo-Calc Equilibrium Simulation |
| $\sigma_{YS} > 200$ MPa at 1300°C | Maresca-Curtin Model **[45]** |
| $\Delta T < 200$ K | Thermo-Calc Equilibrium Simulation |
| CTE < 2% | Thermo-Calc Equilibrium Simulation |
| Single Phase BCC (1300°C to Solidus) | Thermo-Calc Equilibrium Simulation |

*Experimental Methods:* The alloys used for verification were synthesized by arc melting of the elemental metals. The castings were flipped at least three times for improved homogenization. The alloys were then sectioned, mounted, and polished for Rockwell hardness indentation. The indentations were made using a LECO LR-series Rockwell Type hardness tester using a spheroconical diamond tip with 60 kgf (Rockwell A scale) and 150 kgf (Rockwell C scale) load. The indents were imaged using an optical microscope.

**Results and Discussion**

The ductile-to-brittle transition or ductility, in general, is a popular research focus in the context of improving ductility in bcc RMPEAs that precludes their deployment in technological application. This poses a key question: What should be a generalized approach to quantify and improve the ductility in RMPEAs? Pugh [**48**], Lewandowski, et al. [**67**], and many others, including Rice-Thomson [**55**], Zhu et al. [**68**], Hirsch-Roberts [**69**], Hirsch et al. [**70**], Rice [**53**], Cleri et al. [**71**] have provided empirical relations to demarcate ductility with brittleness by studying large classes of crystalline materials. However, most efforts have specialized their examinations to material types in particular stress states, as ductile materials allow massive dislocation emissions and flow. Recently, the dislocation behavior [**72-74**] controlling strength in bcc RMPEAs [**75,76**] was found to inevitably be influenced by local lattice distortions. Unlike in pure metals, RMPEAs have an intrinsic lattice distortion due to mismatch in modulus and atomic sizes that results in activation of multiple dislocation pathways [**77-79**]. This arises from the energetics associated with the local chemical environments due to chemical complexity, which also affects the migration barriers and vacancy formation energies and is correlated with electronegativity differences [**62**]. Lattice distortion can induce local elastic-strain [**80**] that hinders dislocation motion during deformation. Recently, Yang et al. [**81**] has shown that local lattice distortion could significantly alter the dislocation core structure and related Peierls stress in refractory RMPEAs, which is correlated to strength [**82**]. These findings suggest a correlation between local lattice distortion and strength/ductility in bcc RMPEAs. There have been numerous studies connecting strength with local distortion in MPEAs; however, no attempts have been made to develop models or criteria to predict ductility that include the effect of local lattice distortions. To address this gap, in this work, we propose a dimensionless metric of quantum-mechanical origin that uses scalar/vector atomic displacements, including the charge-effect extracted from fully relaxed random supercells using DFT methods. These parameters are a building block for our surrogate model created from positional imbalance around ideal-site symmetries. Furthermore, the *subset of local structural parameters* (i.e., average atomic displacement ($\Delta u$) and vector ($L_{2,1}$) norm of atomic displacements $\left[\sqrt{[\Delta u_{x,y,z}]^2}\,\right]$ from all atoms in supercell) derived from increased charge sharing, and lattice relaxation increases bond strength and, thereby, the mechanical properties. The dimensionless LLD metric is defined as the ratio of average local lattice displacement ($\Delta u_{x,y,z}$) and vector norm of lattice displacements $\left[\sqrt{[\Delta u_{x,y,z}]]^2}\,\right]$ for relaxed supercells, i.e.,

$$LLD = \Delta w_{VEC} \times \frac{\Delta u_{x,y,z}}{\sqrt{[\Delta u_{x,y,z}]^2}} \quad , \qquad (2)$$

where

$$\begin{cases} \text{LLD} < 0.3 \text{ (ductile)} \\ \text{LLD} \geq 0.3 \text{ (brittle)} \end{cases} \qquad (3)$$

and $\Delta w_{VEC}$ (=$VEC_{MPEA}^{bcc}$ - $[VEC_{max}^{bcc} - VEC_{min}^{bcc}]$) is the weighted VEC, i.e., the difference between the VEC of RMPEA with respect to max $VEC_{max}$ and min $VEC_{min}$ values. Here, $VEC_{min}$ and $VEC_{max}$ define the bcc phase formation range based on VEC, i.e., 4 and 6, respectively. The weighted average of VEC difference of RMPEAs in Eq. (2) was defined to remove the superficial dominance of atomic displacement.

The proposed "LLD" metric is significantly different from the "local lattice distortion" calculated using atomic size mismatch, **δ**, in HEAs that is generally adopted from the work of Zhang et al., [**83**], estimated via the relationship **δ** $=100*\sqrt{c_i(1-r_i/\bar{r})^2}$, where $c_i$, $r_i$ and $\bar{r}\ (=\sum_i c_i r_i)$ are concentration, atomic radius of the individual elemental components, and average atomic radii. Unfortunately, this metric is somewhat ambiguous, as the same element in different alloys can show varying atomic radii depending on the crystal structure and the (local) chemical environment [**84-86**]. Moreover, there are several different definitions of atomic radii [**87,88**]. Thus, we redefined the local lattice distortion in terms of an average atomic displacement ($\Delta u$), and vector ($L_{2,1}$) norm of atomic displacements coming from a relaxed supercell.

The $\Delta u_{x,y,z}$ is the distance between displaced atoms from high-symmetry points between relaxed and ideal (average lattice) atomic positions in the supercell due to local lattice mismatch, and $\sqrt{[\Delta u_{x,y,z}]^2}$ is a mean-value derived from vector displacements of all atoms in relaxed supercell. The final ductility metric was calculated as a weighted average of atomic displacements evaluated from the optimal-disordered structures. A new metric is necessary because atoms in various chemical environments can have different effective atomic radii. Therefore, an explicit and comprehensive description of the local lattice distortions (LLDs) is crucial for accurately assessing the ductility of materials. This idea is like Ye et al.'s empirical residual strain (ε) metric for single-phase formation in high-entropy alloys [**89**], where they described three ε ranges, i.e., ε<5% (single phase), 5%<ε<10% (mixed phase), and ε>10% (amorphous phase) for high-entropy alloys. However, the LLD metric in Eq. (2) is more quantum-mechanically-rooted, where the change in electronic effects (charge transfer) is directly considered by including local atomic displacement from DFT optimization.

While bcc RMPEAs are promising candidates for next-generation structural materials owing to their exceptional mechanical properties, but, unfortunately, they often exhibit limited ductility, hindering their broader applications. Consequently, developing a reliable metric for high-throughput screening of useful compositions within the RMPEAs domain would significantly facilitate the identification and optimization of these materials. Therefore, we chose bcc-refractory (ternary, quaternary, quinary) RMPEAs, such as Senkov alloys [**6**], and compared them with the recent work of Curtin et al. [**37**]. We generated a large enough supercell for each case, where the volume (lattice constant) and atomic (co-coordinates) positions were fully optimized. We plot the “LLD” metric with respect to the average atomic displacement (**Fig. 2a**), and $\Delta w_{VEC}$ (**Fig. 2b**) of RMPEAs.

While LLD is by definition linear in $\Delta u_{x,y,z}$, it is renormalized by $\Delta w_{VEC}$ and $\sqrt{[\Delta u_{x,y,z}]^2}$ to remove dimensional bias. This physically interpretable dimensionless metric can greatly simplify the alloy design task. The linear correlation between $\Delta u_{x,y,z}$ and LLD (**Fig. 2a**) shows a clear separation at 0.30 LLD (along y-axis) between experimentally known ductile and brittle behavior, as tabulated in **Table 2**. A clear range in VEC ($\Delta w_{VEC}$=3), $\Delta u_{x,y,z}$ (0.05 Å), and LLD (0.3) is seen in **Fig. 2a,b**, although not as evident for static displacement, $\Delta u_{x,y,z}$=0.05 Å. This suggests that rather than severe lattice distortion, the dimensionless LLD metric, including the effect of electron count, is more physical in characterizing the ductility of RMPEAs, mainly caused by the charge-transfer effect on the mean-value of vector displacements. Furthermore, we plot LLD with respect to DFT-calculated shear moduli for RMPEAs in **Table 2**, with a good correlation LLD trends.

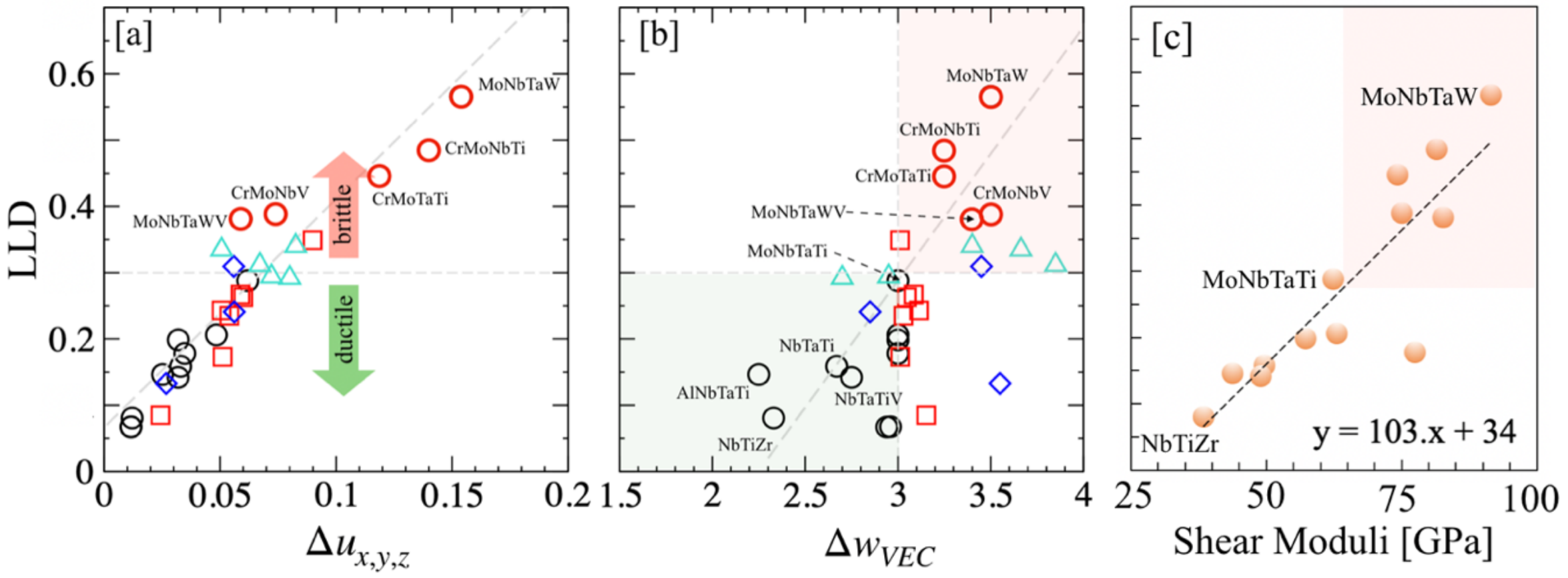

**Figure 2**. LLD metric versus (a) magnitude of vector atomic displacement from the average lattice $\Delta u_{x,y,z}$, and (b) $\Delta w_{VEC}$ [regions: ductile (green) and brittle (red)], and (c) shear moduli (GPa). Ductile behavior of 17 RMPEAs designed in our ULTIMATE program [**65**] was also assessed: (red squares; 9), ( blue diamonds; 3), and ( cyan triangles; 5).

The higher $\Delta u_{x,y,z}$ in **Fig. 2a** shows an inverse correlation with ductility. In literature, LLD is connected to the strength of alloys, but this does not guarantee ductility, especially with bcc RMPEAs that are mostly brittle. Moreover, larger LLD in distorted refractory lattices is expected to induce large local elastic-stress fields. The interaction of mobile dislocations with the local stress fields may hinder dislocation glide during deformation, which accounts for large dislocation density during plastic deformation [**90**]. Recently, Lee *et al.* [**91**] highlighted this by comparing the change in dislocations density near the {110}- and {200}-oriented planes in as-cast versus homogenized RMPEAs. Thus, when characterizing strength, one must consider a reasonable trade-off between ductility and strength when designing alloys using demanding computational schemes. However, the LLD-predicted ductility in **Table 2** shows excellent agreement with other theories and experiments of known bcc RMPEAs [**92**].

**Table 2**. The proposed LLD metric (Eq. 2) to characterize ductility in bcc refractory metals. DFT-derived LLD predictions agree with mean-field theory (MFT) [**37**], experiments [**6,93-98**], and Rice-Thomson (R-T) criteria [**55**], which is compared with tensile elongation of known RMPEAs [**23,99-101**]. Empirical values of lattice mismatch (**$\delta$**) are also provided for comparison.

| **MPEAs** | **$\delta$** | **$\Delta$u** | **$\sqrt{[\Delta u]^2}$** | **$\Delta w_{VEC}$** | **LLD** | **R-T** | **$\varepsilon_t$ (%)** | **Curtin** | **this work** | |
|---|---|---|---|---|---|---|---|---|---|---|
| NbTiZr | 4.22 | 0.012 | 0.351 | 2.33 | 0.081 | 29.2 | 14.2 | 1 | 3 | |
| AlNbTaTi | 1.03 | 0.025 | 0.387 | 2.25 | 0.146 | 28.7 | -- | 2 | 7 | |
| $Nb_{6.8}Mo_{1.4}Ti_{1.8}$ | 2.59 | 0.012 | 0.506 | 2.94 | 0.067 | -- | -- | 3 | 1 | |
| $Nb_{7.0}Mo_{1.2}Ti_{1.8}$ | 2.62 | 0.012 | 0.506 | 2.96 | 0.068 | -- | -- | 4 | 2 | |
| NbTaTi | 0.32 | 0.033 | 0.555 | 2.67 | 0.159 | 33.6 | 18.2 | 5 | 5 | |
| NbTaV | 3.98 | 0.048 | 0.600 | 3 | 0.243 | 30.4 | -- | -- | new | |
| NbTaTiV | 3.74 | 0.032 | 0.617 | 2.75 | 0.142 | 31.4 | 11.8 | 6 | 4 | **Ductile** |
| MoNbTiV | 3.76 | 0.032 | 0.483 | 3 | 0.199 | 36.56 | 25 | 7 | 8 | |
| MoNbTaTi | 2.22 | 0.062 | 0.645 | 3 | 0.288 | 40.8 | -- | 8 | 10 | |
| MoNbTaV | 3.59 | 0.048 | 0.704 | 3 | 0.206 | 37.8 | 21 | 9 | 9 | |
| NbTaTiW | 2.22 | 0.035 | 0.587 | 3 | 0.178 | 50.1 | -- | 10 | 6 | |
| | | | | | | | | | | **LLD=0.3** |
| CrMoTaTi | 5.42 | 0.140 | 0.939 | 3.25 | 0.484 | 50.9 | -- | 11 | 14 | |
| CrMoNbTi | 5.42 | 0.119 | 0.866 | 3.25 | 0.446 | 46.8 | -- | 12 | 13 | **Brittle** |
| CrMoNbV | 4.83 | 0.074 | 0.666 | 3.5 | 0.388 | 43.2 | 4.2 | 13 | 11 | |
| MoNbTaWV | 3.28 | 0.059 | 0.527 | 3.4 | 0.381 | 49.1 | 1.7 | 14 | 12 | |
| NbTaMoW | 2.46 | 0.154 | 0.950 | 3.5 | 0.566 | 55.4 | 2 | 15 | 15 | |

The use of lattice-distortion parameters in determining the local atomic displacement for RMPEAs is somewhat ambiguous. This possibly relates to the definition of 'local' in relaxed atomic supercells and its correlation with charge (calculated from DFT) or electronic configuration (valence electron count) with atomic sizes. However, we found a clear correlation between static displacement and effective electronic charges with LLD metric, see **Fig. 2a**. We found that the high LLD is induced mainly by the atomic-size mismatch in RMPEAs, which will reduce the deformability of alloy by hindering slip plane easy glide [**91**]. **Fig. 2a** shows that the LLD metric [Eq. 2 & 3] gives a reliable ductility (deformability) prediction.

The VEC has been connected to the deformability in alloys, where a lower VEC (<4.5) is expected to improve the ductility. We plot $\Delta w_{VEC}$ vs. LLD metric in **Fig. 2b** and found that atomic displacement has more to do with atomic sizes in RMPEAs than the charge of constituent elements or electron configuration. With each RMPEA marked in **Fig. 2b**, the displacement or LLD metric increases with alloys having increasing at.% of elements with increasing VEC, such as Cr, Mo, or W. Along with a clear demarcation in LLD (=0.30), we also found a clear value in $\Delta VEC$ that separates the ductile behavior of RMPEAs ($\Delta w_{VEC} < 3.0$) from brittleness ($\Delta w_{VEC} > 3.0$). To add to the choice of LLD metric cut-off, higher LLD plays a key role in impeding the dislocation motion in RMPEAs, which changes the deformation mechanism and reduces ductility [**74,75**]. The severe lattice distortion is expected to lessen the crystallite growth rate, causing amorphous structures to form. Thus, the choice of LLD range in Eq. (3) seems reasonable and agrees with the experimental tensile elongation as tabulated in **Table. 2**.

We have shown in **Fig. 2b** that ductile RMPEAs fall in the area shaded in green, where Al/Ti/Zr are revealed as the main elements that drive bcc ductility, in agreement with previous work [**102,103**]. The alloy will be brittle if the elastic instability mode transitions from tensile failure to shear failure after reaching the ideal tensile stress. The comparison between LLD and tensile elongation in **Table 2** reflects this fact. Furthermore, lowering the VEC will increase the driving force for the Jahn-Teller distortions, which results in earlier shear instability and lowers the total energy of the alloy, i.e., increasing ductility.

Our LLD metric predictions are in good agreement with the recent work of Curtin et al. [**44**], where NbTaMoW (#14) and MoNbTaWV (#15) were predicted to be brittle with less than 3% ductility at RT under compression, while NbTiZr (#1) was predicted to be ductile. These predictions agree with existing experiments that show ductility for NbTaZr [**102,103**] and brittleness for NbTaMoW and MoNbTaWV [**6**]. Furthermore, the LLD metric assessed the ductility of 7 new quinary Ti-V-Nb-Mo-W RMPEAs (Nb-rich, i.e., Nb >60 at.%), shown in **Fig. 2** (red circles). Six RMPEAs satisfy the LLD metric constraint out of seven, i.e., LLD < 0.3 in Eq. (3), but only two of them ($Mo_{1.5}Nb_{74}V_{23}W_{1.5}$ and $Mo_{1.1}Nb_{68.4}Ti_{1.5}V_{27.4}W_{1.6}$) satisfy critical LLD and ΔVEC limit for an RMPEA to be ductile. We also show the empirical lattice distortion parameter (**δ**) in **Table 2** arising from size-mismatch calculated using atomic radii of elements. However, no such correlation with ductility was observed in contrast to the LLD metric. We attribute this fact to the inclusion of the quantum-mechanical charge effect in the LLD through atomic displacements in Eq. (2,3).

Recently, Geslin et al. have pointed out the effect of finite unit-cell size on displacement and local stress [**104,105**]. As our goal was to evaluate quantities related to displacements directly required by LLD metric, Eq. (2), we investigated the finite cell-size effect on local displacements. Specifically, we examined the LLD in quaternary NbTaTiV as a function of supercell size. For that, we tested SCRAPs for number of atoms (bcc supercell size) [16, 32, 60, 72, 128, and 160 atoms per supercell] with LLD values of [0.578; 0.264; 0.142; 0.153; 0.137; 0.141], respectively. We assessed spatial correlation or cell size on LLD. Therefore, we attribute these changes to large displacement "$\Delta u$," which is nearly {0.151/0.719 (16); 0.109/0.976 (60)} for smaller supercells while the large supercells cells show {0.037/0.735 (128) to 0.051/0.915 (72)}. This confirms the findings of Geslin et al [**104,105**] regarding finite cell-size effects. However, in our context, this effect was found to be small beyond 60 atom unit cells for stoichiometric compositions.

In **Figure 2a,b**, DFT-calculated LLD for 17 RMPEAs belonging to the Mo-W-Ti-V-W (7), Mo-Nb-Ta-V-W (3), and Cr-Mo-Nb-V-W (5) family. These alloys were designed from high-throughput CALPHAD and analytical models to satisfy high-temperature strength criteria (>300 MPa and 1300°C). Our analysis shows that 4 out of 15 RMPEAs meet all ductility criteria, i.e., LLD < 0 .3, 2 < $\Delta w_{VEC}$ < 3, and $\Delta u_{x,y,z}$ < 0.1, and 6 satisfy LLD and Δu criteria but fall above ΔVEC range. In contrast, the remaining 5 RMPEAs were found to be brittle, i.e., they do not meet any of the criteria set for ductility. Nevertheless, the main advantage of the "LLD" metric is our ability to calculate it in a high-throughput manner using first-principles DFT methods. Therefore, LLD is appropriate for screening regions with improved ductility from the vast space of RMPEAs.

We also assessed the ductility of RMPEAs in **Table 2** using the Rice-Thomson criterion, where the authors considered the actual dislocation processes due to the localized nature of shear at the crack tip [**48**]. The Rice-Thomson criterion quantifies the stability of a sharp crack against the emission of a blunting dislocation in a crystal. Therefore, qualitatively the crystals with wide core dislocations or small values of *Gb*/γ are ductile. In contrast, alloys with large values of *Gb*/γ (or narrow dislocation cores) are brittle, where *b, G,* and *γ* are the Burger's vector, shear modulus, and surface energy, respectively. In **Fig. 3a**, we compared LLD and Rice-Thomson criterion, which shows a good trend with observed ductility and brittle behavior of RMPEAs in **Table 2**. The Rice-Thomson criterion [**55**] was also assessed with respect to VEC (**Fig. 3b**) and $\Delta u_{x,y,z}$ (**Fig. 3c**), which shows a good correlation with experimental trends. Our findings suggest that the local lattice distortion (LLD) metric provides a more accurate assessment of ductility near weakly ductile regions, such as NbTaTiW, compared to Rice-Thomson criterion, which classifies these regions as brittle. Although, we found good linear trend with Rice-Thomson model, but it fails to capture correct Gb/γ range, which is expected to be < 7.5-10 for ductile materials while > 10 for brittle. However, LLD consistently captures the prospered ductility vs brittle range in Eq. (2,3).
Additionally, calculating surface energy is computationally expensive, while evaluating the LLD metric is simpler. Therefore, knowledge of the LLD metric offers a reliable and efficient predictive tool for guiding the discovery of new alloys with improved ductility.

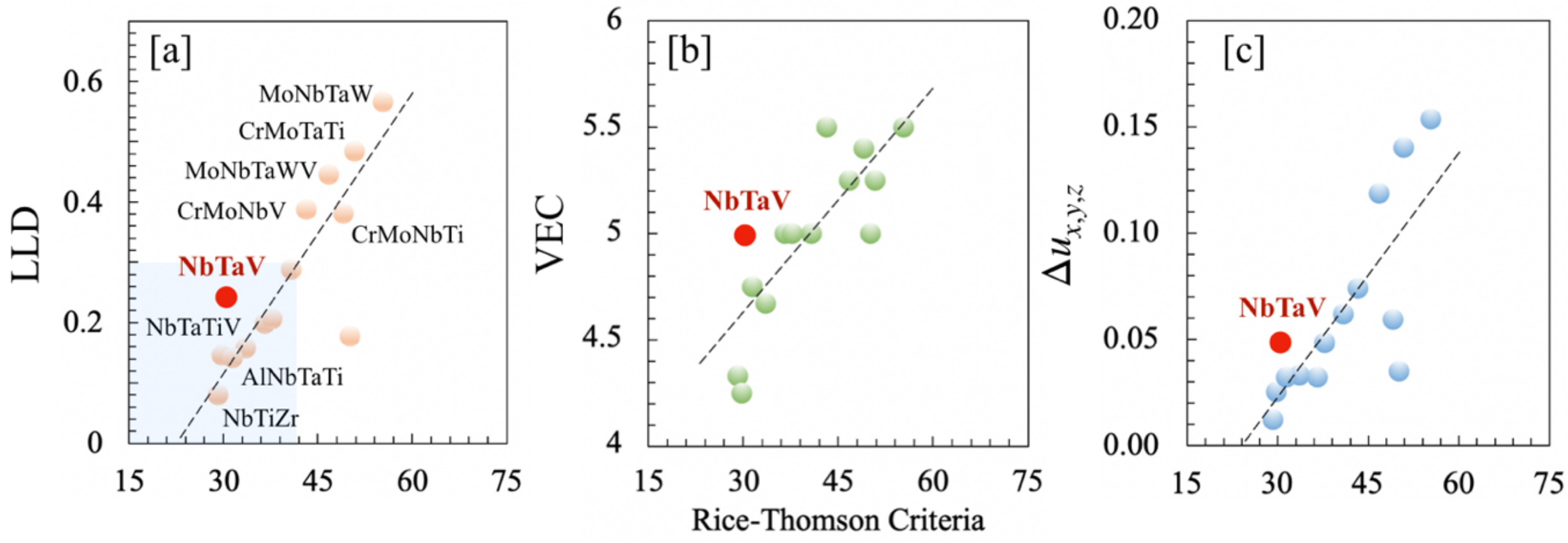

**Figure 3**. Comparison DFT-calculated Rice-Thomson criteria [**53**] with (a) LLD, (b) VEC, and (c) $\Delta u_{x,y,z}$. The ductile region is marked based on the criterion shown in Eq. (2).

To validate the ductility prediction, we compared the LLD metric with tensile elongation ($\boldsymbol{\varepsilon_t}$) of known refractory RMPEAs in **Table 2**. The LLD metric range in Eq. (2,3) demarcating ductile vs. brittle behavior was found to show similar trends as found in experiments [**23,99-101**]. Notably, elongation or fracture strain, which is the percentage increase in length that material will achieve before breaking, is very low for NbTaMoW(V) (2%, and 1.7%) and Cr-based alloys (~4.2%) RMPEAs, which consistent with LLD metric.

**Ductility analysis of ternary NbTaV RMPEA**

In **Fig. 3**, we also show the ductility prediction for ternary NbTaV (red circle). The DFT-predicted LLD and Rice-Thomson values are 0.243 and 30.4, respectively, which suggests that the NbTaV should be ductile. This seems reasonable given NbTaTi (LLD=0.159 "<0.3"; R-T=33.6) and NbTaTiV (LLD=0.142 < "0.3"; R-T=31.4) are also experimentally found to be ductile as shown by higher tensile-elongation in **Table 2**. Notably, both LLD and R-T fall in the lower half corner, a ductile zone, in **Fig. 2**. We further analyzed NbTaV to understand the origin of ductility. In **Fig. 4a**, we show a fully-relaxed supercell (54 atom/cell) showing polyhedral distortion near V sites as marked by P1 and P2. The marked regions in **Fig. 4b** show increased charge activity, attributed to higher electronegativity for V (1.62) on the Allen scale for solids compared to Nb (1.17) and Ta (1.30). For clarity, in **Fig. 4c**, we have shown $\Delta\rho$ projected on (101) plane containing P1 and P2 from **Fig. 4b** (full projection is shown in **Fig. A3**). This suggests that higher $\chi$ affects $\Delta u_{x,y,z}$ that affects the charge-transfer ability of other elements.

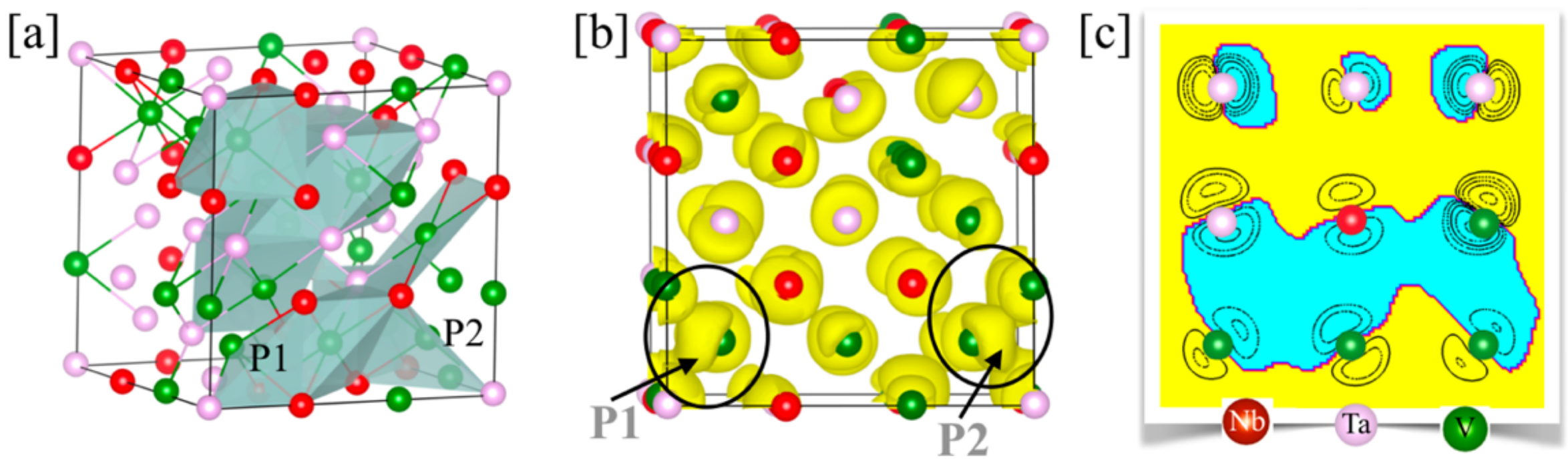

**Figure 4**. (a) 54-atom supercell for ternary NbTaV, (b) charge-density difference ($\Delta\rho$ [LLD – no-LLD]), and (c) 2D project $\Delta\rho$ on [100] plane showing charge activity at (P1, P2) in (b).

Hu et al. [**106**] have shown that elemental radii do not accurately characterize lattice distortion from a change in the local environment, as distortions (and related properties) are strongly dependent on the local chemical environment in RMPEAs [**61,107**]. Moreover, the electronegativity difference of elements was found to correlate well with the mechanical properties [**108**], rather than the Hume-Rothery size effect [**109**] and data-driven approaches [**110**]. To understand the effect of chemical complexity (alloying and environments) on local lattice distortion, we compare in **Fig. 5,** for three fully-relaxed quaternary alloys AlNbTaTi, VNbTaTi, and CrMoNbTi, the relaxed SCRAP structures (**Fig. 5a,d,g**), local atomic displacement ($\Delta u_{x,y,z}$) in **Fig. 5b,e,h**, and the change in local charges from varying neighbor environments (**Fig. 5d,f,i**).

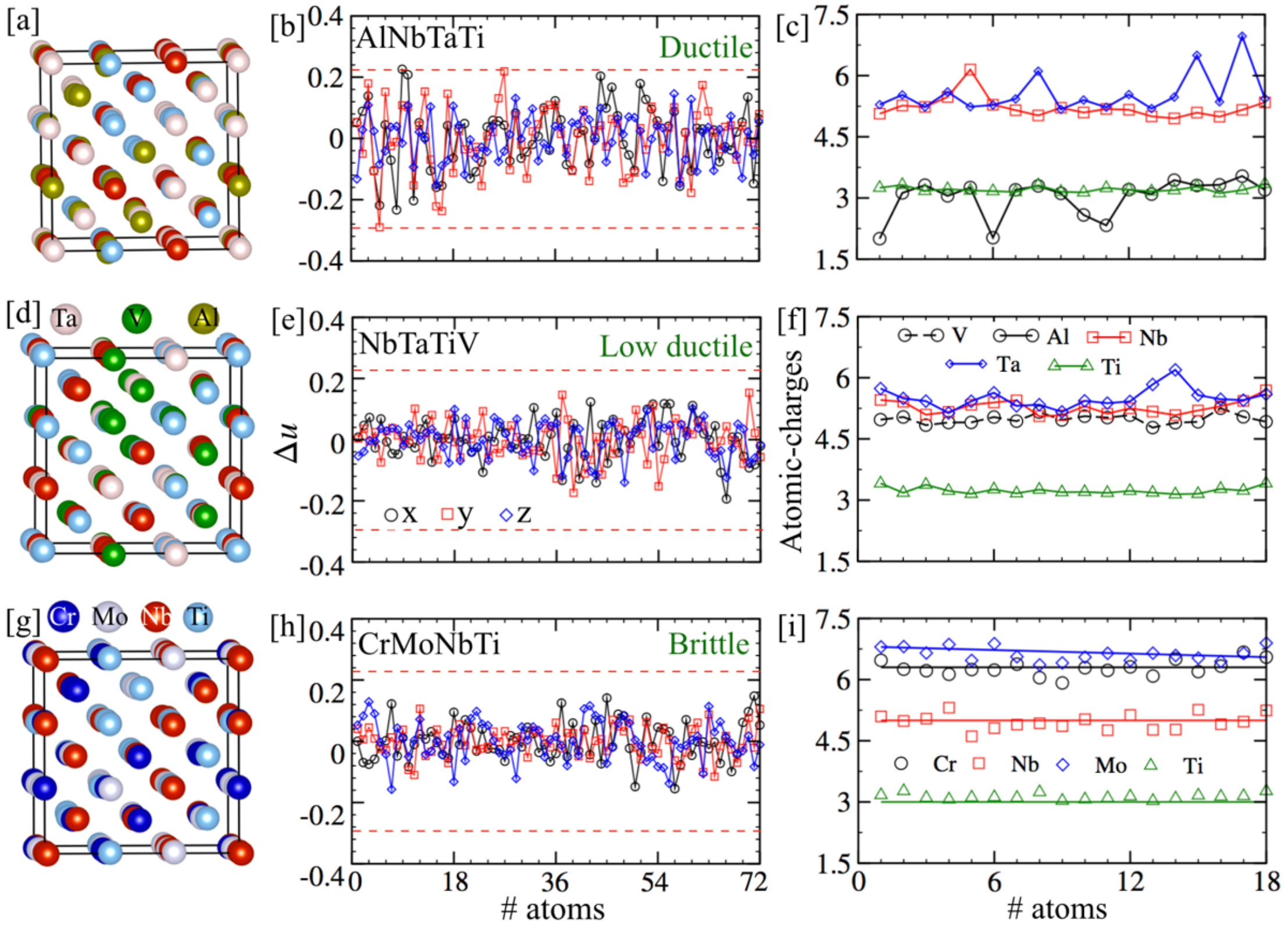

**Figure 5.** Relaxed SCRAPs and plots of local lattice displacement and charge-transfer activity in (a-c) AlNbTaTi (high ductility), (d-f) NbTaTiV (low ductility), and (g-i) CrMoNbTi (no ductility).

**Figure 5b,e,h** illustrates the likelihood of an atomic site to be displaced (compressed or elongated) based on alloying species, their electronegativities, and atomic sizes in AlNbTaTi, NbTaTiV, and CrMoNbTi RMPEAs. The bond-length analysis of Nb and Ti is shown in **Fig. 5** (also in **Fig. A1**). We found clear elongation in the Nb-X bond length around Al or Nb in particular, while Nb-X bonds in CrNnTaTi show compression or no change compared to unrelaxed structures. Similarly, Ti-X bonds in NbTaTiAl and NbTaTiV show weak elongation, while Ti-X in CrNbTaTi shows small reduction or no change in bond length. This Nb-X and Ti-X elongation and compression of bonds are also reflected in volume change in three RMPEAs, where $V_{AlNbTaTi}$ (17.3 Å$^3$/atom) > $V_{NbTaTiV}$ (16.6 Å$^3$/atom) > $V_{CrNoNbTi}$ (15.4 Å$^3$/atom). More specifically, the homoatomic pairs, i.e., Nb-Nb and Ta-Ta, are atoms with large radii. Thus, these are primarily compressed, while the homoatomic pairs of Ti-Ti and V-V are the smaller size atoms with significant elongation in RMPEAs. The idea of bond elongation or compression based on alloying elements and their intrinsic characteristic is reflected through absolute lattice displacement in **Fig. 5b,e,h**. The change

in average bond lengths is also related to enhanced charge transfer due to varying chemical interactions of the principal elements in RMPEAs.

The order of elemental electronegativities ($\chi$) on the Allen scale for solids is Nb (1.17), Ta (1.30), Cr (1.33), Mo (1.38), Ti (1.40), Al (1.52), and V (1.62). So, for NbTaTiAl $\chi_{Nb}<\chi_{Ta}<\chi_{Ti}<\chi_{Al}$, NbTaTiV $\chi_{Nb}<\chi_{Ta}<\chi_{Ti}<\chi_{V}$, and CrMoNbTi ($\chi_{Nb}<\chi_{Cr}<\chi_{Mo}<\chi_{Ti}$). Clearly, CrMoNbTi has the most elements with lower affinities for pulling charge from their neighbors, i.e., the least distortion expected due to charge transfer, as reflected in **Fig. 5c,f,i**. Charge analysis in **Fig. 5c,f,i** indicates both $\chi$ and $\Delta u$ due to atomic size are strongly correlated with charge-transfer ability of alloying elements, as is quantified in LLD metric, which combines distortion parameters ($\Delta u_{x,y,z}$ and $\sqrt{[\Delta u_{x,y,z}]^2}$ ) with change in VEC in electronically distinct RMPEAs.

The relationship between ductility and strength can further be understood from the charge density difference ($\Delta\rho$) between two RMPEAs, revealing the nature of bonding and charge transfer. Thus, for three RMPEAs, we analyzed the $\Delta\rho$ of three RMPEAs that show high-ductility (AlNbTaTi), low-ductility (NbTaTiV), and no-ductility/brittle (CrMoNbTi), see **Fig. 2** and **Table 2**. The $\Delta\rho_{CrMoNbTi\,-\,AlNbTaTi}$ in **Fig. 6a** shows increased charge transfer near Al in AlNbTaTi, while Cr-based CrMoNbTi shows almost no charge activity at/near Cr. The increased charge near/at Al can be attributed to large Al electronegativity ($\chi$ =1.52) compared to Nb- or Cr-based RMPEAs. Also, increased charged activity or metallic interaction introduced by Al arises from delocalized 3s-3p electrons, while the increased charge activity at/near Ta ($4d^3 5s^2$) is due to larger atomic size. This suggests multiple mechanisms that are responsible for improved ductility in RMPEAs. On the other hand, the effective charge transfer between brittle (CrMoNbTi) and weakly ductile (NbTaTiV) RMPEAs in **Fig. 6b** shows improved charge activity, as shown by $\Delta\rho_{CrMoNbTi\,-\,NbTaTiV}$. Similarly, the $\Delta\rho_{NbTaTiV\,-\,AlNbTaTi}$ in **Fig. 6c** shows much larger charge transfer activity suggesting increased lattice distortion for AlNbTaTi and NbTaTiV RMPEAs, as shown in **Fig. 5a-c** and **Fig. 5d-f**, respectively. While CrMoNbTi in **Fig. 5g-i** shows very weak or no charge transfer, as is visible through local lattice distortion, mainly expected to originate from the atomic-size effect.

The bond length in HEAs can be ambiguous as all atoms within a certain separation distance can contribute to metallic bonding. Thus, the multi-atomic nature of metallic bonds makes them different from ionic or covalent bonding. In **Fig. 6d-f**, we plot the average bond-length distribution (see also **Fig. A2**) around each species in high-ductility (AlNbTaTi), low-ductility (NbTaTiV), and brittle (CrMoNbTi) RMPEAs. The resulting average atomic charges and total valence are shown in **Table 3**. Bond lengths can take different values for pairs of atoms due to the complex environment in RMPEAs.

Therefore, average bonding was considered for each atom. Notably, the average nearest-neighbor interatomic distance in most ductile RMPEA, i.e., AlNbTaTi in **Fig. 2d**, was found to be much larger but uniformly distributed chemical bonds. However, the nearest-neighbor interatomic distance for all pairs in the least ductile RMPEAs, i.e., CrMoNbTi RMPEAs, extends over a wide range (**Fig. 6f)** compared to AlNbTaTi (**Fig. 6d**) and NbTaTiV (**Fig. 6e**), where Nb (36-53) and Ti (54-71) atoms show profound deviation from ideal sites that shows increased (severe) lattice distortion. The widely distributed yet nearly 5.4% smaller bond lengths of CrMoNbTi compared AlNbTaTi indicates higher strength [**23**]. Notably, our LLD metric predicts poor ductility for CrMoNbTi, agreeing with experiment [**31**]. Our hypothesis that alloying metal elements of diverse electronegative will improve the ductility (**Table 2**) and correlation with strength (**Fig. A4**) of RMPEAs is consistent with our predictions.

**Table 3.** DFT-calculated effective elemental charges and total valence in a unit cell for quaternary AlNbTaTi, NbTaTiV, and CrMoNbTi RMPEAs**.**

| **RMPEAs** | **Effective elemental charges** | | | | **Total valence** |
|---|---|---|---|---|---|
| **Al-Nb-Ta-Ti** | 3.226 (Al) | 4.870 (Nb) | 4.800 (Ta) | 4.102 (Ti) | 17e- |
| **Nb-Ta-Ti-V** | 4.857 (Nb) | 4.790 (Ta) | 4.094 (Ti) | 5.257 (V) | 19e- |
| **Cr-Mo-Nb-Ti** | 6.324 (Cr) | 5.888 (Mo) | 4.743 (Nb) | 4.043 (Ti) | 21e- |

In **Fig. 6g-i**, we plot the partial density of states for AlNbTaTi, NbTaTiV, and CrMoNbTi to understand electronic-structure changes in ductile vs. brittle RMPEAs. As seen by the PDOS in **Fig. 6g**, it shows an obvious overlap among all alloying elements. The increased overlap indicates strong electron hybridization of Al-*3p* with *3d* and *4d* bands of Ti/Nb/Ta transition metals and an increase in covalency, where flat yet localized conduction bands in the high energy region indicate stronger interaction of orbital electrons [**111,112**]. A thin valence band near the low energy region (-5 eV) below the Fermi-level was found for AlNbTaTi (**Fig. 6g**), which comes entirely from Al-*3p*. The delocalized nature of Al-*3p* leads to the formation of metallic bonding, which is expected to be the electronic-structure reasoning for increased ductility in AlNbTaTi. Similar features (flat bands at low energy, i.e., below Fermi-level, and substantial band overlap) were found in NbTaTiV (**Fig. 6h**) arising from the presence of V instead of Al. However, no such features were observed in CrMoNbTi (**Fig. 6i**), which the LLD metric predicted to be poorly ductile.

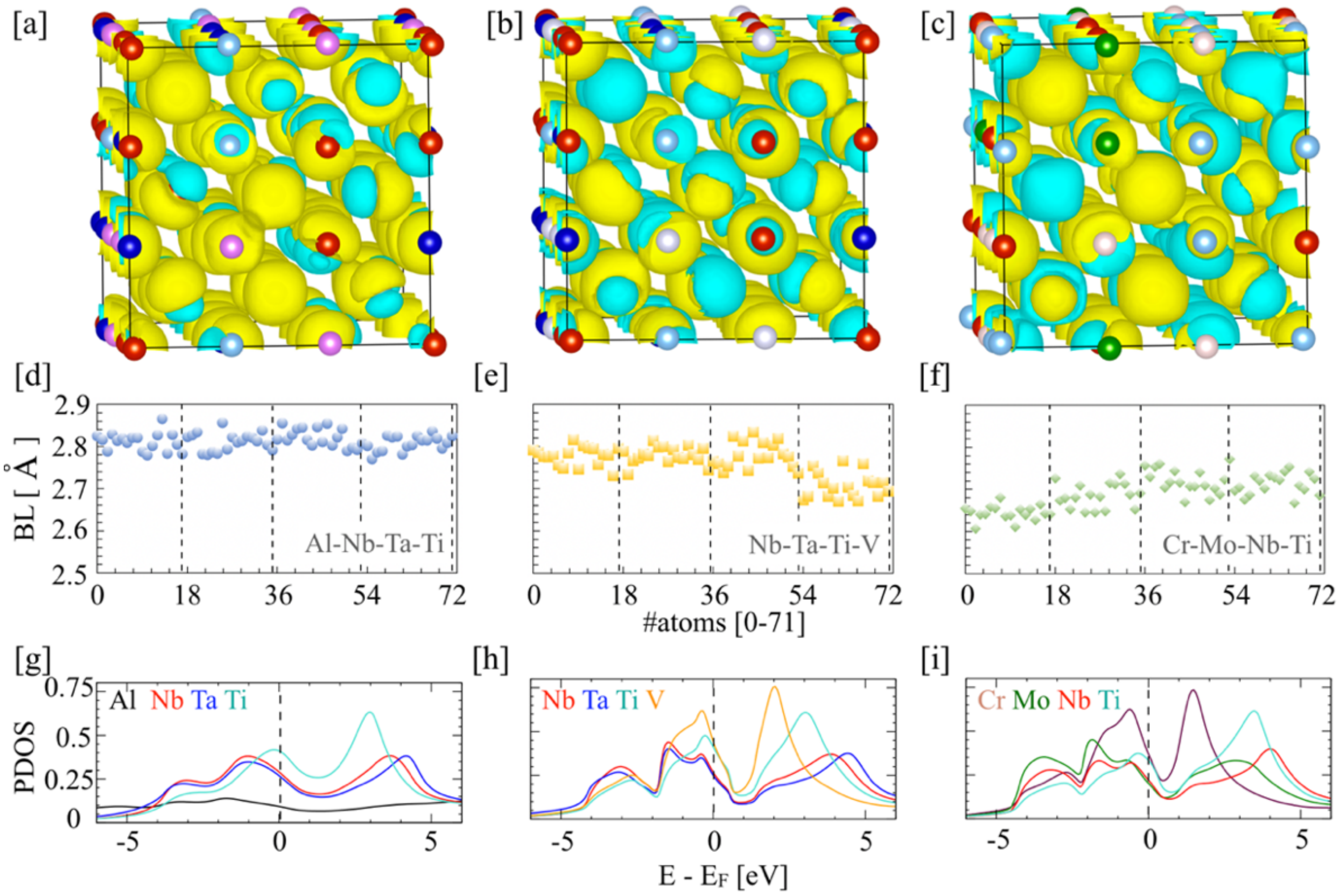

**Figure 6**. Charge-density difference (a) $\Delta\rho$ [CrMoNbTi – AlNbTaTi], (b) $\Delta\rho$ [CrMoNbTi – NbTaTiV], and (c) $\Delta\rho$ [NbTaTiV – AlNbTaTi] to emphasize a change in charge distribution (iso-surface value of 0.012 $e^-/Å^3$). Iso-surfaces are positive (yellow) and negative (blue) charges. (d-f) Bond-length distribution about each element in fully relaxed 72-atom SCRAP (A [0-17]; B [18-35]; C [36-53]; D [54-71]). Partial density of states (PDOS) for (g) AlNbTaTi, (h) NbTaTiV, and (i) CrMoNbTi RMPEAs.

## LLD prediction and experimental validation

We chose two MPEAs viz NbTaMoW and $Mo_{72}W_{13}Ta_{10}Ti_{2.5}Zr_{2.5}$ to test and validate our metric prediction. The LLD metric predicts brittleness for NbTaMoW (LLD=0.566 > 0.3, **Eq. 3**) and ductility for $Mo_{72}W_{13}Ta_{10}Ti_{2.5}Zr_{2.5}$ (LLD=0.255 < 0.3, **Eq. 3**). To qualitatively evaluate the ductility of two RMPEAs, we performed indentation tests using Rockwell indents at different force and observe the crack formation through optical microscopy as shown in **Fig. 7**.

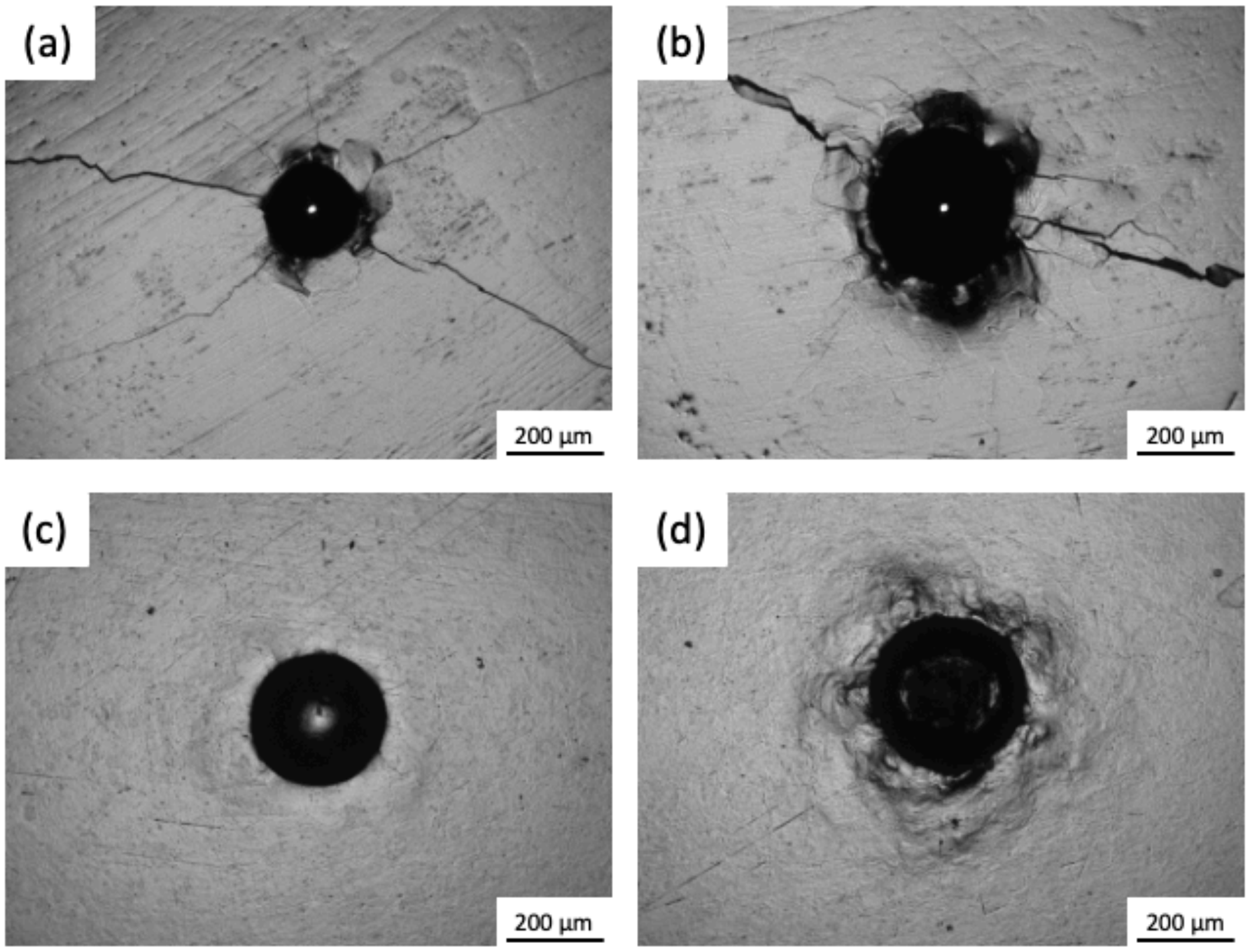

**Figure 7**. Optical images around Rockwell indents showing crack formations on (a,b) NbTaMoW (LLD=0.566) but not on (c,d) $Mo_{72}W_{13}Ta_{10}Ti_{2.5}Zr_{2.5}$ (LLD=0.255). This suggests improved ductility for $Mo_{72}W_{13}Ta_{10}Ti_{2.5}Zr_{2.5}$, as predicted by LLD metric in **Table 4**. The indents are made using Rockwell hardness testers with 60 kgf (a,c) and 150 kgf (b,d) load, respectively.

Under both 60 kgf major load and 150 kgf major load conditions, multiple cracks formed around the indent on the NbTaMoW sample (**Fig. 7a,b**). Larger load has resulted in broader crack formation. In contrast, the minimum crack formation was observed on the $Mo_{72}W_{13}Ta_{10}Ti_{2.5}Zr_{2.5}$ sample under both conditions (**Fig. 7c,d**). The surface impression around the indent in **Fig. 7d** attains that $Mo_{72}W_{13}Ta_{10}Ti_{2.5}Zr_{2.5}$ RMPEA can sustain a great level of plastic deformation without crack formation. The crack length and indentation size can also be used to compute fracture toughness, pioneered by Evans and Charles [**114**].

A practical formula is Nihara's method, as in ASTM silicon-nitride bearing balls standard F2094.

$$K_{IFR} = 10.4\,(E^{0.4})(P^{0.6})(a^{0.8}/c^{1.5}), \quad (4)$$

where $K_{IFR}$, E, P, a, c is the indentation fracture resistance, elastic modulus, applied load, mean half-diagonal length, and mean half-tip-to-tip crack length, respectively. Though the exact values calculated by this method may bear some error [**115**], the significantly reduced crack length in $Mo_{72}W_{13}Ta_{10}Ti_{2.5}Zr_{2.5}$ RMPEA suggests its much-improved fracture toughness than NbTaMoW RMPEA. This agrees with our compression test results [**116,117**], showing improved compressive ductility for $Mo_{72}W_{13}Ta_{10}Ti_{2.5}Zr_{2.5}$. The findings from both indentation and compression tests are consistent with our ductility prediction based on the LLD metric.

**Comparison of LLD predictions with other ductility models**

To showcase the improvements in ductility prediction, we present comparative study of LLD predictions for NbTaMoW and $Mo_{72}W_{13}Ta_{10}Ti_{2.5}Zr_{2.5}$ with other commonly used ductility models such as VEC, Cauchy pressure, Pugh's ratio (PR), and D parameter. The VEC is an easy to evaluate and empirical metric used for fast assessment of ductility in alloys. Recently, Sheikh et al. [**113**] suggested that single-phase refractory RMPEAs made from elements of groups 4, 5, and 6 should be ductile for VEC $< 4.5$ and brittle when VEC $\geq 4.5$. Based on VEC criteria, both MPEAs in **Table 4** are expected to be brittle, which is not true as shown by indentation experiments in **Fig. 7**. The Cauchy pressure is based on the idea of angular covalent bonding to characterize ductility. If Cauchy pressure ($C_{12}$-$C_{44} < 0$) is negative the alloy will be brittle and ductile for positive pressure ($C_{12}$-$C_{44} > 0$). A very high positive Cauchy pressure for both RMPEAs in **Table 4** suggests ductility, which is contrary to both LLD prediction and experiments. Pugh's ratio (PR) is the most widely used metric. Pugh suggested that if the ratio of bulk moduli vs shear moduli (B/G) greater than 1.87 the alloy is expected to be ductile. Based on PR metric values in **Table 4**, both MPEAs have significantly higher value of Pugh's ration than 1.87, therefore, expected to be ductile. However, the prediction of Pugh's ratio is contrary to both LLD and experiments. The D-parameter is another metric proposed by Rice [**53**] that uses dislocation emission flow in alloys to predict ductility. The low positive values of D parameter in **Table 4** suggest high brittleness or low fracture toughness as discussed by Hu et al [**52**]. To summarize, no ductility metric other than LLD was able to distinguish right ductility behavior for two MPEAs in **Table 4** (other examples area discussed in **Table 2**).

**Table 4**. LLD-predicted ductility (more details in **Table A2**) for NbTaMoW and $Mo_{72}W_{13}Ta_{10}Ti_{2.5}Zr_{2.5}$ MPEAs, and comparison with ductility models from literature determine VEC, Cauchy pressure, Pugh's ratio (PR), and D parameter (Rice). Note that LLD correctly identifies the ductility and brittleness behavior in contrast to other models.

| | **MPEAs** | **Metric** | **Ductile or Brittle?** | **Observed** |
|---|---|---|---|---|
| **This work** | | | | |
| **LLD** | NbTaMoW | 0.566 | Brittle | Yes |
| | $Mo_{72}W_{13}Ta_{10}Ti_{2.5}Zr_{2.5}$ | 0.255 | Ductile | Yes |
| **Literature** | | | | |
| **VEC** | NbTaMoW | 5.5 | Brittle | Yes |
| | $Mo_{72}W_{13}Ta_{10}Ti_{2.5}Zr_{2.5}$ | 5.8 | Brittle | No |
| **Cauchy** | NbTaMoW | 87.2 | Ductile | No |
| | $Mo_{72}W_{13}Ta_{10}Ti_{2.5}Zr_{2.5}$ | 57.5 | Ductile | Yes |
| **Pugh's ratio** | NbTaMoW | 2.65 | Ductile | No |
| | $Mo_{72}W_{13}Ta_{10}Ti_{2.5}Zr_{2.5}$ | 2.58 | Ductile | Yes |
| **D parameter** | NbTaMoW | 2.1 | Brittle | Yes |
| | $Mo_{72}W_{13}Ta_{10}Ti_{2.5}Zr_{2.5}$ | 2.5 | Brittle | No |

### LLD vs. other ductility criteria

Several criteria or models exist that try to predict ductility or plastic behavior in alloys [**29,118,119**]. The Pugh ratio [**47**] and Cauchy pressure [**120**] are by far the most widely used due to availability of straight forward ways for measuring or calculating elastic parameters, especially bulk (B), shear (G), and $C_{12}$-$C_{44}$ elastic moduli. Notably, both models exclude the direct consideration of crystal structure and change in local geometry. On the other hand, the proposed LLD in Eq. (2) includes effects arising from changes in the crystal structure, local geometry (atomic lattice), and chemistry (composition) through supercell consideration. Furthermore, Pugh ratio and Cauchy pressure criteria ignore the dislocation mobility [**121**], which would not change Pugh ratio but alter the yield stress. These limitations make Pugh ratio and Cauchy pressure criteria unfavorable in providing accurate predictions of ductility in novel materials. In contrast, dislocation emissions-based criteria can better assess the ductility in metal alloys. The Rice-Thompson [**55**], Rice [**53**], and Zhou-Carlsson-Thomson [**122**] are the three most used models that consider the emission of dislocations from a sharp crack tip to characterize ductility. Instead of fracture and dislocation glide used in Pugh and Cauchy criteria, the fracture and dislocation nucleation form the basis of dislocation-emission-based models that require the energy barrier experienced by dislocations, usually characterized by $\gamma_{us}$ or $b \times G$, where $\gamma_{us}$ is the unstable stacking fault energy. The dislocation emission criteria tend to be more accurate than Pugh determine Cauchy but are computationally expensive and have poor transferability. In contrast, the LLD metric is computationally inexpensive requiring a single run of supercell relaxation. Moreover, LLD shows the potential to overcome drawbacks of elasticity (Pugh ratio and Cauchy pressure) and dislocation-emission (Rice-Thompson, Rice, and Zhou-Carlsson-Thomson models) based on ductility criteria, which has been elaborated in **Table 4** through direct comparison among various ductility models presented.

**Computational difficulty of LLD vs dislocation-emission-based models**

The local-lattice distortion calculations require only a single relaxation run of disordered supercell to arrive at the LLD value in **Eq. (2,3)**. On the other hand, dislocation-emission-based models requires surface energy (gamma in the Rice-Thomson or Rice model) that necessitates multiple calculations including design of orientation-dependent disorder supercell, for example, {110} surface for bcc. Additionally, the reorientation of the already large, disordered supercell must be increased by 2x-3x to conserve the parent MPEA composition. Lastly, the calculation of surface energy involves full relaxation followed by self-consistent runs to determine energies in different configurations, which contrasts with single step LLD metric. Thus, LLD metric is easier to implement and calculate in high-throughput manner compared to dislocation-emission-based models.

**Conclusion**

We provided a detailed understanding of how electronic (charge-transfer) effects affect the local lattice distortion in bcc RMPEAs, and correlate local lattice distortion to ductility. The peculiar characteristics of RMPEAs have produced several design strategies to achieve strength-ductility synergy [**64,123-127**], for use in advanced structural applications, requiring high strength and high ductility. The characterization of alloys purely based on lattice distortion may give an idea about higher strength but not ductility. Importantly, alloys with low ductility are typically not useful for technological applications. Therefore, a strength-ductility trade-off must be utilized. The proposed LLD metric and limits for bcc RMPEAs show that higher lattice distortion leads to poor ductility. As is well-known, bcc RMPEAs have higher strength and generally poor ductility. Presently, our analysis provides valuable guidelines for optimizing LLD and strength to achieve a sweet spot for strength and ductility. As we have shown, ductility in RMPEAs is strongly correlated to local charge-transfer activity and lattice distortion, which can be tuned by alloying. The charge-transfer activity, electronic structure, bond lengths, and lattice distortion for MPEAs were determined from DFT calculations. Our results provide a more fundamental understanding of role charge transfer plays in controlling local lattice distortion and ductility of RMPEAs.

In conventional alloys with a regular lattice, dislocation movements need to overcome the Peierls friction or the lattice stress through a kink-pair mechanism [**128**]. However, increased lattice distortion and the resultant residual stress field in RMPEAs may enable significant strengthening [**129,130**], improving their yield strength. Our study suggests, however, that increased lattice distortion is not necessarily good for ductility. The good combination of strength and ductility derives from increased lattice friction yielding mechanical features beyond those reported before for bcc alloys. The proposed metric successfully rationalized the ductility across a range of ternary, quaternary, and quinary RMPEAs. The LLD metric-driven analysis is validated by comparing it with tensile elongation of available experiments, which establishes the accuracy of identifying ductility behavior. Finally, the proposed LLD metric will contribute to optimizing ductility in refractory-based alloys to accelerate novel refractory RMPEA development [**131**].

**Acknowledgments**

Authors acknowledge the U.S. Department of Energy (DOE) ARPA-E ULTIMATE Program through Project DE-AR0001427. BV acknowledges the support of NSF through Grant no. 1746932. RA also acknowledges NSF through Grant No. 2119103. The part of LLD analysis for BCC alloys was supported by Laboratory Directed Research and Development Program (LDRD) program at Ames Laboratory. The work at Ames Laboratory was supported by the U.S. Department of Energy (DOE) Office of Science, Basic Energy Sciences, Materials Science & Engineering Division. The research was performed at Iowa State University and Ames Laboratory, which is operated by ISU for the U.S. DOE under contract DE-AC02-07CH11358.

**Author Contributions**

P.S. – Conceptualization, Method development, Data curation, Formal Analysis, Writing – original draft, Writing – review and editing. B.V. – Data curation, Formal Analysis, Writing – original draft, Writing – review and editing. G.O. – Data curation, Formal analysis, Writing – original draft, Writing – review and editing. N.A. – Formal analysis, Writing – review and editing. G.C. – Supervision, Writing – review and editing. R.A.– Supervision, Formal analysis, Writing – review, and editing. D.D.J.– Supervision, Formal analysis, Writing – review and editing.

**Appendix:**

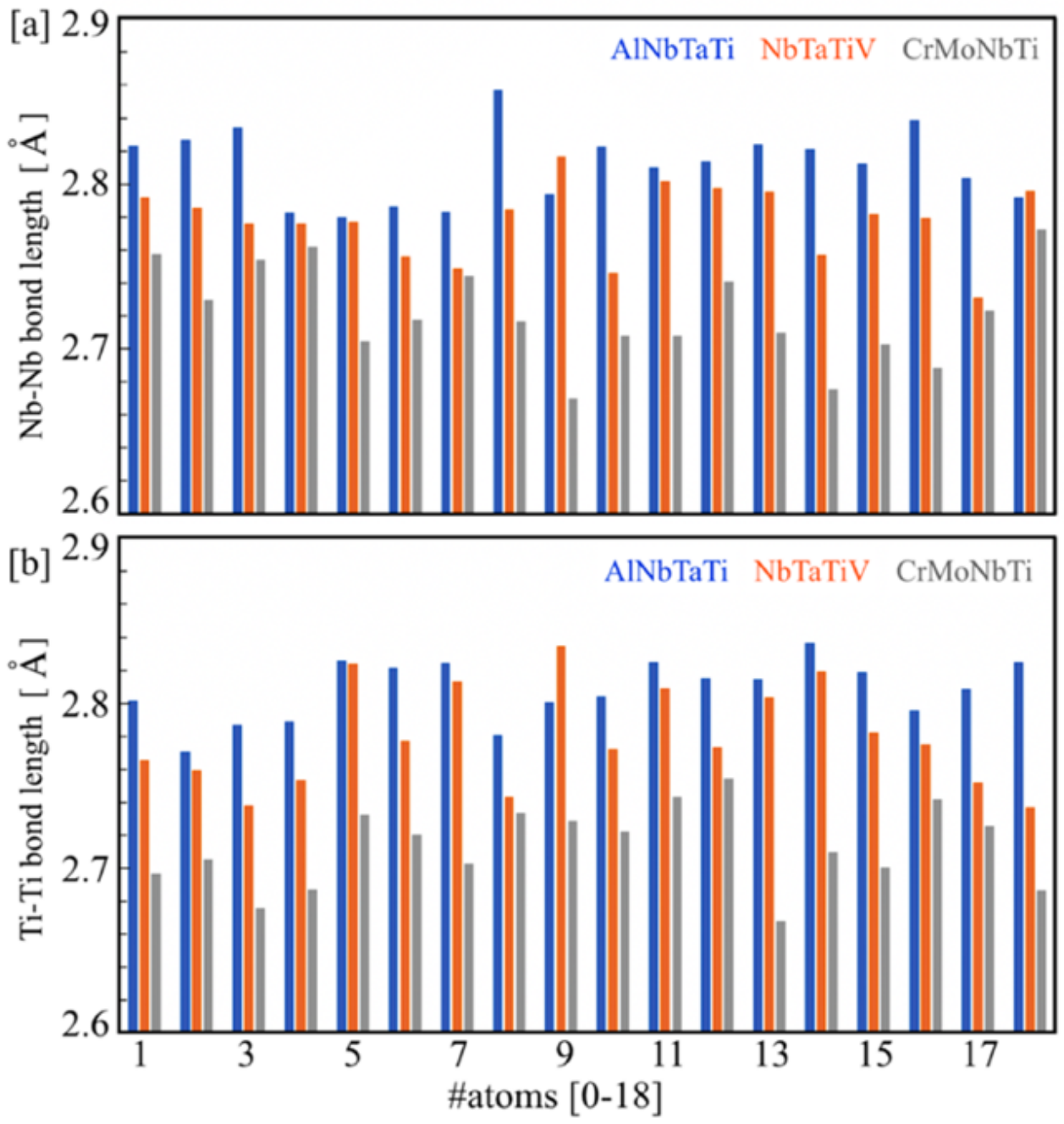

**Figure A1**. (a) Nb-Nb, and (b) Ti-Ti bond-length compression and elongation in (a) AlNbTaTi, (b) NbTaTiV, and (c) CrMoNbTi due to change in chemical environment in RMPEAs.

In **Fig. A2**, we also show the bond-length distribution of selected RMPEAs, where AlNbTaTi with higher ductility shows smooth (Gaussian) bond-distribution compared to skewed (bimodal) distribution in medium and low ductility RMPEAs (here NbTaTiV and CrMoNbTi).

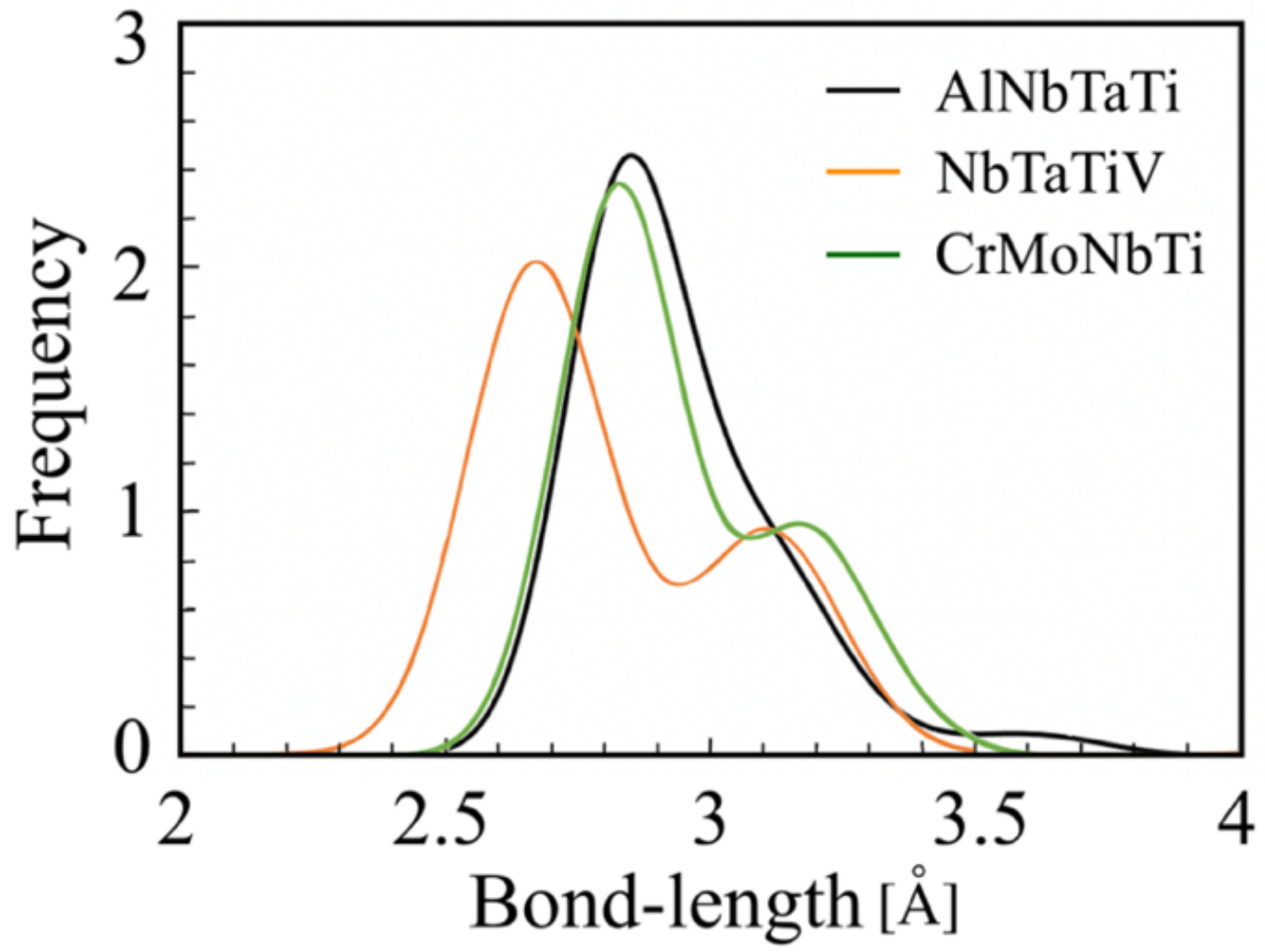

**Figure A2**. The distribution of atomic bond-lengths, and (b) the minimum distance between the atomic surface from the maximum cut-off radius for the core region versus local atomic volume.

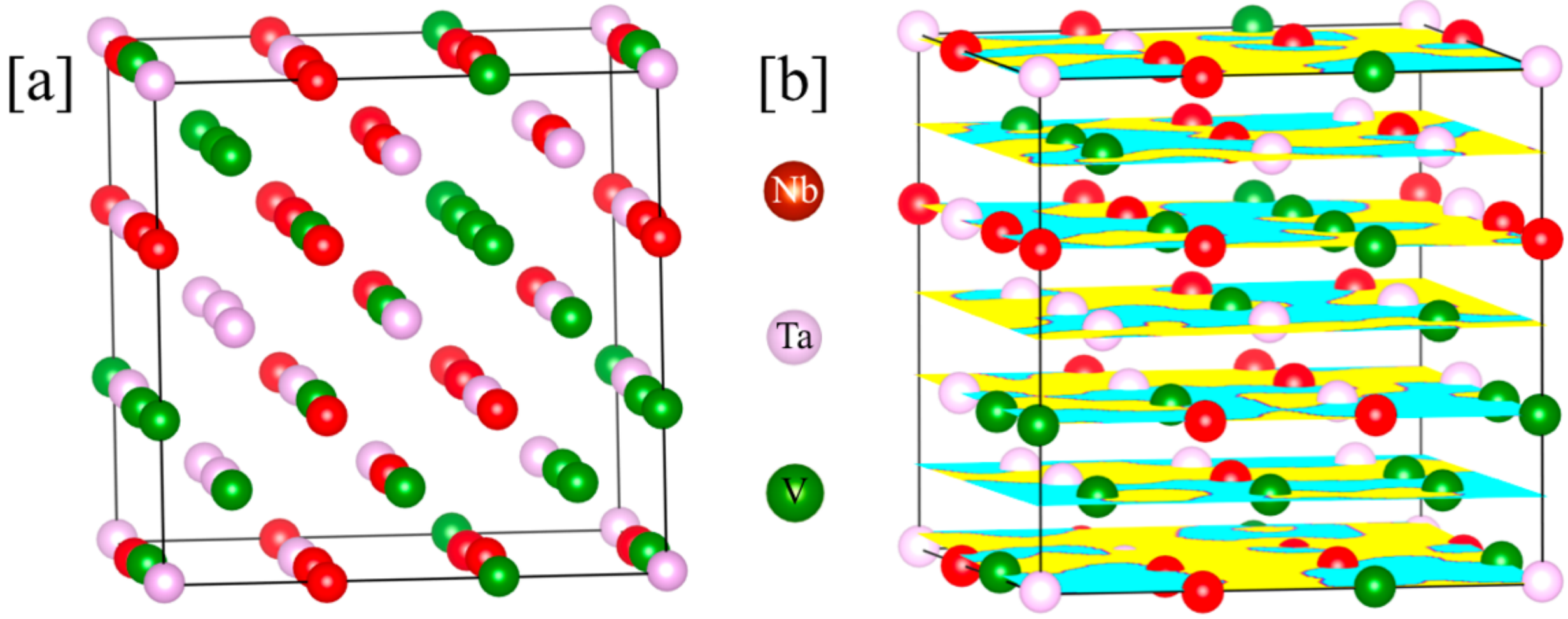

**Figure A3**. (a) A 54-atom supercell of fully relaxed ternary bcc NbTaV. (b) 2D projected charge density difference calculated at same lattice constants for fully relaxed vs. non-relaxed supercells.

**Comment on strength-ductility relationship in MPEAs**: The strength and ductility tradeoff in materials can be obtained after thermal treatment and/or thermo-mechanical processing by controlling the microstructure to remove defects. To understand this tradeoff using the LLD metric, we plot in **Fig. A4** the LLD and average atomic size for MPEAs in **Table 2** concerning experimental yield strength (YS) [**132**]. The trend between LLD and YS in **Fig. A4a** shows an increase in strength with LLD up to 0.3, which then decreases with an increase in LLD. This suggests that maximizing lattice distortion is not always a reason for increased strength, i.e., there is an optimal LLD range for tuning strength and ductility tradeoff in MPEAs. Furthermore, in **Fig. A4b**, the plot of YS as a function of atomic size shows a linear correlation, such that atomic size increases the strength in bcc MPEAs from associated larger lattice distortions, which also suggests that the increment of YS can be predicted.

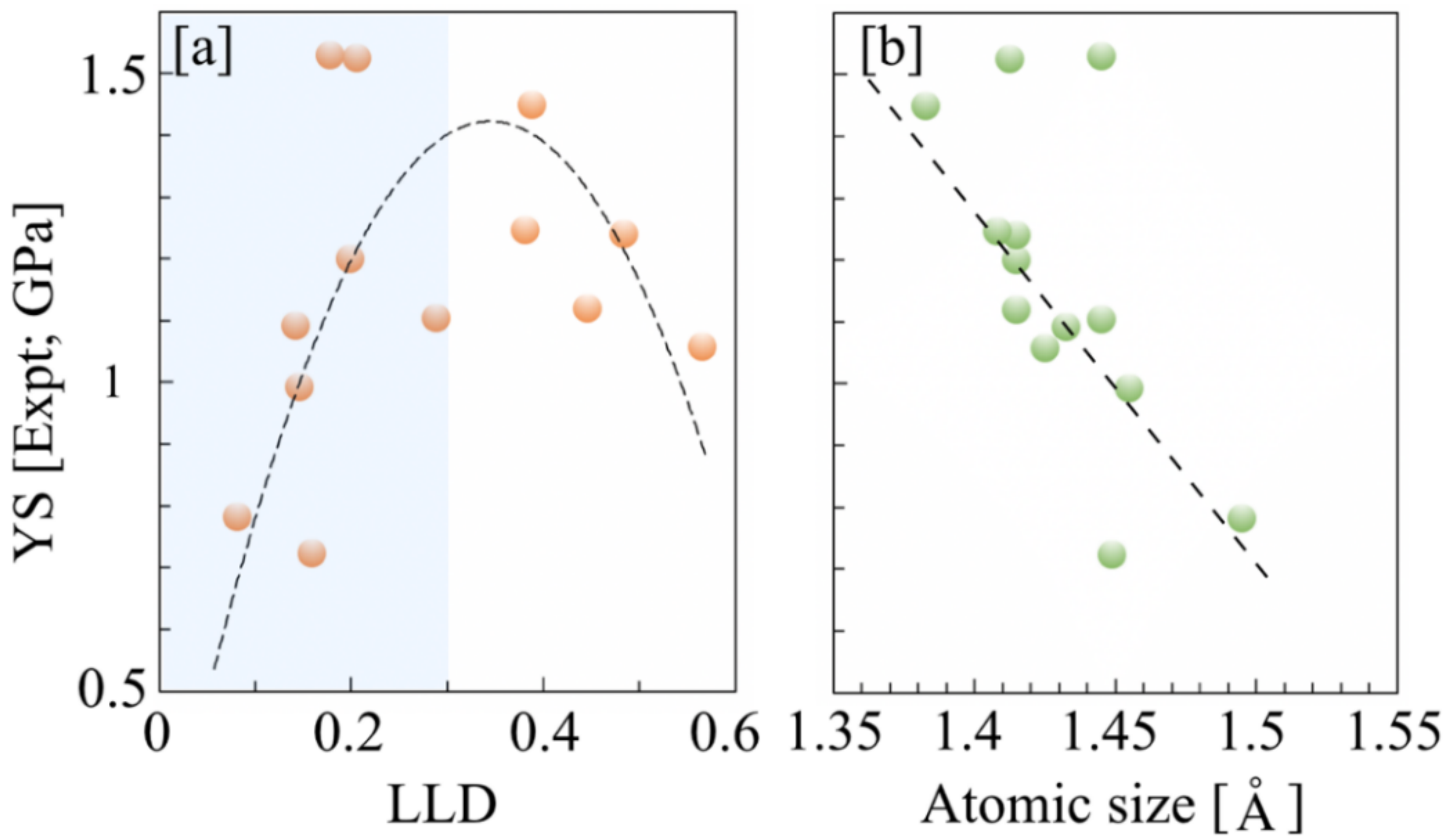

**Figure A4**. Measured yield strength (GPa) for bcc MPEAs vs. LLD (a) and atomic size (b). Higher ductility connects with a lower strength (see **Fig. 2**), shaded area in (a). MPEA strength decreases with increasing average atomic size (b).

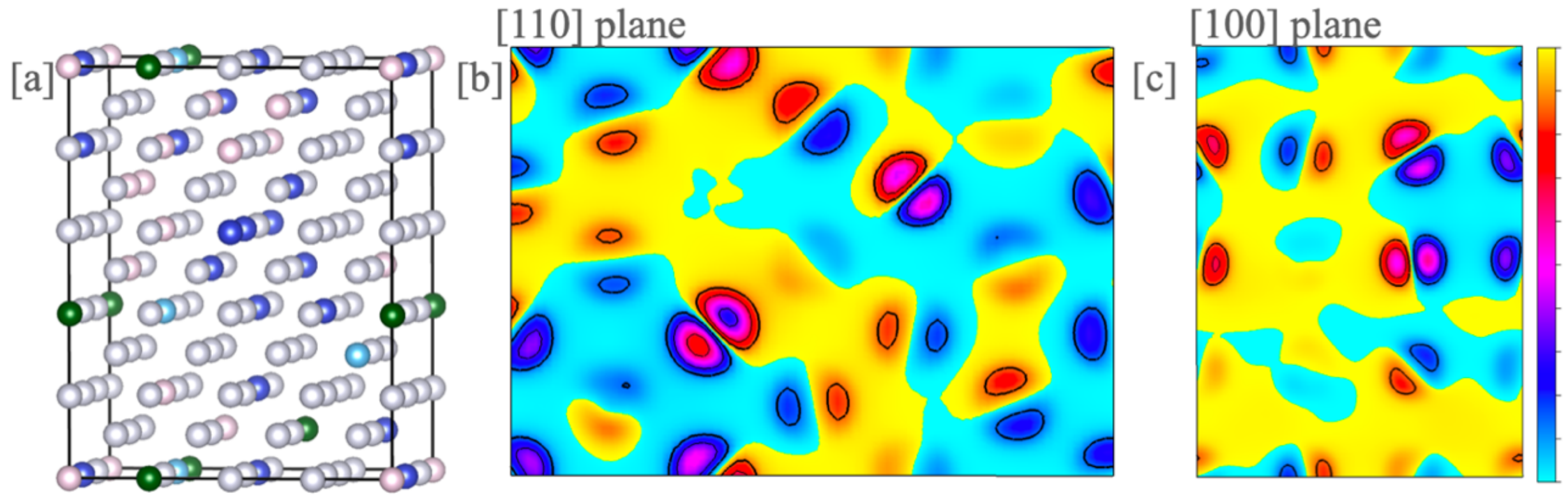

**Figure A5**. (a) Crystal structure of 120 atom quinary BCC $Mo_{72}W_{13}Ta_{10}Ti_{2.5}Zr_{2.5}$ MPEA, and increased charge activity along (b) (110), and (c) (100) planes.

**Table A1.** The fracture strain data (compressive) and ductility metrices including D-parameter (Rice) [**52**] as well as rule-of-mixture Pugh Ratio (B/G), Cauchy pressure ($C_{12}$- $C_{44}$) and valence electron count (VEC) for 56 MPEAs in **Figure 1**.

| **MPEAs** | **Fracture Strain (%)** | **D parameter** | **Pugh Ratio** | **Cauchy Pressure** | **VEC** |
|---|---|---|---|---|---|
| $TiZrV_{0.3}Nb$ | 45 | 3.485 | 3.11 | 92.02 | 4.39 |
| $TiZrV_{0.3}NbMo_{0.1}$ | 45 | 3.387 | 3.08 | 91.24 | 4.44 |
| $TiZrV_{0.3}NbMo_{0.5}$ | 43 | 3.106 | 2.98 | 88.56 | 4.61 |
| $TiZrVNbMo_{0.3}$ | 42 | 3.185 | 3.10 | 95.91 | 4.60 |
| ZrHfNbTa | 34 | 3.316 | 3.30 | 87.50 | 4.50 |
| TiZrNbMo | 33 | 2.883 | 2.84 | 82.80 | 4.75 |
| $TiZrVNbMo_{0.5}$ | 32 | 3.085 | 3.05 | 94.57 | 4.67 |
| $TiZrVNbMo_{0.7}$ | 32 | 3 | 3.01 | 93.34 | 4.72 |
| TiZrVNbMo | 32 | 2.89 | 2.96 | 91.68 | 4.80 |
| $TiZrV_{0.25}NbMo$ | 30 | 2.893 | 2.87 | 85.41 | 4.76 |
| $TiZrVNbMo_{1.3}$ | 30 | 2.795 | 2.91 | 90.21 | 4.87 |
| TiVNbTaMo | 30 | 2.809 | 2.96 | 97.82 | 5.00 |
| TiZrHfVNb | 29.6 | 3.569 | 3.24 | 91.60 | 4.40 |
| $TiZrV_{0.75}NbMo$ | 29 | 2.896 | 2.93 | 89.81 | 4.79 |
| $Ti_{1.5}ZrHfNbMo$ | 28.98 | 3.177 | 2.90 | 78.96 | 4.55 |
| $TiZrV_{0.5}NbMo$ | 28 | 2.898 | 2.90 | 87.73 | 4.78 |
| $TiZrV_{0.3}NbMo_{0.7}$ | 26.6 | 3.011 | 2.93 | 87.41 | 4.68 |
| TiZrVNbMo | 26 | 2.89 | 2.96 | 91.68 | 4.80 |
| TiVNbMo | 25.62 | 2.728 | 3.00 | 98.58 | 5.00 |
| $TiZrV_{0.3}NbM$ | 25 | 2.894 | 2.88 | 85.90 | 4.77 |
| $TiZrHf_{0.5}NbMo_{0.5}$ | 24.61 | 3.237 | 3.00 | 82.75 | 4.50 |
| $TiZrHf_{0.5}NbMo_{0.5}$ | 24.61 | 2.997 | 3.00 | 82.75 | 4.50 |
| $TiZrV_3NbMo$ | 24 | 2.874 | 3.10 | 101.83 | 4.86 |
| $TiZrHfNb_{1.5}Mo$ | 23.97 | 3.058 | 3.06 | 83.53 | 4.64 |
| $TiZrV_2NbMo$ | 23 | 2.877 | 3.04 | 97.60 | 4.83 |
| $TiZrV_{1.5}NbMo$ | 20 | 2.881 | 3.00 | 94.91 | 4.82 |
| TiVNbTaW | 20 | 2.827 | 2.83 | 97.14 | 5.00 |
| $TiZrV_{0.3}NbMo_{1.3}$ | 20 | 2.793 | 2.83 | 84.58 | 4.85 |
| $TiZr_{0.5}HfNbMo$ | 18.02 | 3.038 | 2.98 | 80.99 | 4.67 |
| $TiZrHf_{1.5}NbMo$ | 16.83 | 3.162 | 3.00 | 78.03 | 4.55 |
| $TiZr_{1.5}HfNbMo$ | 16.09 | 3.135 | 2.94 | 77.92 | 4.55 |
| TiNbTaMoW | 14.1 | 2.498 | 2.61 | 84.84 | 5.20 |
| $TiZrHfNb_{0.5}Mo$ | 13.02 | 3.13 | 2.84 | 74.13 | 4.56 |
| $TiZrHf_{0.5}NbMo$ | 12.09 | 2.997 | 2.90 | 80.86 | 4.67 |
| $Ti_{0.5}ZrHfNbMo$ | 12.08 | 2.979 | 3.02 | 79.71 | 4.67 |
| TiZrHfNbTaMo | 12 | 3.073 | 2.93 | 81.88 | 4.67 |
| NbTaVW | 12 | 2.527 | 2.95 | 102.53 | 5.25 |
| TiNbTaMoW | 11.5 | 2.498 | 2.61 | 84.84 | 5.20 |
| $TiZrHfNbMo_{1.5}$ | 10.83 | 2.924 | 2.88 | 78.06 | 4.73 |
| TiVNbTaMoW | 10.6 | 2.575 | 2.73 | 91.90 | 5.17 |
| TiZrHfNbMo | 10.2 | 3.088 | 2.96 | 79.30 | 4.60 |
| TiZrHfNbMo | 10.12 | 3.088 | 2.96 | 79.30 | 4.60 |
| VNbTaMoW | 8.8 | 2.28 | 2.79 | 95.16 | 5.40 |
| $Ti_{0.75}NbTaMoW$ | 8.4 | 2.424 | 2.62 | 85.33 | 5.26 |
| $TiZrV_{0.3}NbMo_{1.5}$ | 8 | 2.731 | 2.81 | 83.79 | 4.90 |
| $Ti_{0.5}NbTaMoW$ | 5.9 | 2.335 | 2.63 | 85.87 | 5.33 |
| NbTaMoW | 2.6 | 2.113 | 2.65 | 87.15 | 5.50 |
| $Ti_{0.25}NbTaMoW$ | 2.5 | 2.233 | 2.64 | 86.47 | 5.41 |
| NbTaMoW | 2.1 | 2.113 | 2.65 | 87.15 | 5.50 |
| NbTaMoW | 1.9 | 2.113 | 2.65 | 87.15 | 5.50 |
| VNbTaMoW | 1.7 | 2.28 | 2.79 | 95.16 | 5.40 |
| VNbTaMoW | 1.7 | 2.28 | 2.79 | 95.16 | 5.40 |

**Table A2**. LLD-predicted ductility for NbTaMoW and $Mo_{72}W_{13}Ta_{10}Ti_{2.5}Zr_{2.5}$ RMPEAs, and comparison with currently used metrices in literature including VEC, Cauchy, Pugh's ratio (PR), and D parameter.

| MPEAs | δ | Δu | $\sqrt{[\Delta u]^2}$ | $\Delta w_{VEC}$ | LLD | VEC | Cauchy | Pugh | D | |
|---|---|---|---|---|---|---|---|---|---|---|
| NbTaMoW | 2.46 | 0.154 | 0.950 | 3.5 | 0.566 | 5.5 | 87.2 | 2.65 | 2.1 | Brittle |
| $Mo_{72}W_{13}Ta_{10}Ti_{2.5}Zr_{2.5}$ | 2.80 | 0.029 | 0.431 | 3.8 | 0.255 | 5.8 | 57.5 | 2.58 | 2.5 | Ductile |